\documentclass[nofootinbib]{article}
\usepackage{graphicx}
\usepackage[hang, nooneline]{subfigure} 
\usepackage[utf8]{inputenc} 
\usepackage{amsmath} 
\usepackage{amssymb}
\usepackage{bm}
\usepackage{color} 
\usepackage{url}

\usepackage{tikz}
\usetikzlibrary{shapes,arrows}

\begin{document}

\title{An adaptive selective frequency damping method}

\author{Bastien     E.     Jordi, Colin  J.   Cotter, Spencer  J. Sherwin}     
\date{\today}


\maketitle

\begin{abstract}
The selective frequency damping (SFD) method is an alternative to classical Newton's method to obtain unstable steady-state solutions of dynamical systems. However this method has two main limitations: it does not converge for arbitrary control parameters; and when it does converge, the time necessary to reach the steady-state solution may be very long. In this paper we present an adaptive algorithm to address these two issues. We show that by evaluating the dominant eigenvalue of a ``partially converged'' steady flow, we can select a control coefficient and a filter width that ensure an optimum convergence of the SFD method. We apply this adaptive method to several classical test cases of computational fluid dynamics and we show that a steady-state solution can be obtained without any \textit{a priori} knowledge of the flow stability properties.
\end{abstract}

\section{Introduction}

To numerically compute stability analysis with high accuracy, it is crucial to carefully choose the base flow around which the governing equations will be linearised. The steady-state solution is mathematically appropriate because it is a solution of the system considered. In computational fluid dynamics, if a flow is linearly stable, obtaining a steady-state solution is trivial, we only have to execute the code and wait long enough until the flow becomes constant in time. However for unstable flows, obtaining a steady-state solution is a concrete challenge. The problem of finding an unstable fixed point of a non-linear system has to be addressed. 

The selective frequency damping (SFD) method \cite{Steady_NS_SFD} appears to be an efficient alternative to classical Newton's method to solve this problem in the field of fluid dynamics. It is based on the filtering of unstable temporal frequencies. In its encapsulated formulation \cite{Encaps_SFD}, the SFD method is very easy to implement as a wrapper function around an existing unsteady solver. This method can only be applied if the dominant unstable eigenvalue (\textit{i.e.} the one with the largest modulus) of the flow has an imaginary part which is non-zero. If not, infinitesimal perturbations of the base flow have a pure exponential growth and no frequency can be damped by the method. Hence the SFD method is not an appropriate tool to find the steady-state solution of wall confined jets \cite{sherwin2005three}, for example. However the SFD method can be used to obtain the steady-state solutions of flows with oscillatory growth of the instabilities. 

The convergence of the SFD method is governed by two parameters, which are the control coefficient $\chi$ and the filter width $\Delta$. For arbitrary choice of these parameters, the method may not be able to control the evolution of the instabilities within the flow. Hence the method does not always converge. Even when a steady-state solution can be found, convergence may be very slow. The selection of the parameters $\chi$ and $\Delta$ is central for users of the SFD method. We intend to address this issue in this study. 

We present an adaptive procedure that couples the SFD method and a global stability analysis method. The idea is to approximate the dominant eigenvalue of the flow studied during the execution of the solver implementing the SFD method. This approximation is used to tune $\chi$ and $\Delta$ by using a simple one-dimensional model. The strength of this procedure is that it does not require any knowledge of the flow behaviour before executing the code.

In Sec. \ref{SectionEncapsSFD} we summarise the properties of the encapsulated formulation of the SFD method. In Sec. \ref{SectionStabAnal} we present the main aspects of global stability analysis and recall a modified Arnoldi iteration method \cite{ModifiedArnoldi} that evaluates the dominant eigenvalue of a given base flow. In Sec. \ref{SectionOptParam} we show how optimum parameters of the SFD method can be selected when the dominant eigenvalue of the system studied  is known. In Sec. \ref{SectionAdaptive} we present an adaptive procedure to automatically select parameters that ensure an optimum convergence of the SFD method. Finally in Sec. \ref{SectionNumSim} we show that this procedure can be successfully applied to obtain unstable steady-state solutions of several classical test cases of computational fluid dynamics. This set of test cases aims to validate the adaptive method prior to its application to more challenging flows.


\section{Encapsulated SFD method}  
\label{SectionEncapsSFD}

In this section we recall theoretical aspects of the encapsulated formulation of the SFD method. With appropriate initial and boundary conditions, any dynamical system can be written
\begin{equation}
\dot{\bm{\mbox{q}}}=\bm{\mbox{F}}(\bm{\mbox{q}}),
\label{system}
\end{equation}
where $\bm{\mbox{q}}$ represents the problem unknown(s), the dot represents the time derivative and $\bm{\mbox{F}}$ is an operator (which can be nonlinear). The steady-state $\bm{\mbox{q}}_s$ of this problem is reached when $\dot{\bm{\mbox{q}}}_s=\bm{\mbox{F}}(\bm{\mbox{q}}_s)=0$.

The time continuous formulation of the SFD method is 
\begin{equation}
\begin{cases}
\dot{\bm{\mbox{q}}}=\bm{\mbox{F}}(\bm{\mbox{q}})-\chi (\bm{\mbox{q}}-\bar{\bm{\mbox{q}}}), \\
\dot{\bar{\bm{\mbox{q}}}}=\frac{\bm{\mbox{q}}-\bar{\bm{\mbox{q}}}}{\Delta},
\end{cases}
\label{SFD-General}
\end{equation} 
where $\chi$ is the control coefficient (real and positive), $\bar{\bm{\mbox{q}}}$ is a filtered version of $\bm{\mbox{q}}$, and $\Delta$ is the filter width of a first-order low-pass time filter (real and strictly positive). The steady-state solution is reached when $\bm{\mbox{q}}=\bar{\bm{\mbox{q}}}$.

For a given space and time discretization of (\ref{system}), we define $\bm{\bm{\Phi}}$ as the function reading $\bm{\mbox{q}}^{n+1}$ and $\bm{\mbox{q}}^{n+1}$ \textit{i.e.}
\begin{equation}
\bm{\mbox{q}}^{n+1} = \bm{\Phi} (\bm{\mbox{q}}^n).
\label{Subproblem1}
\end{equation}

To implement the encapsulated formulation of the SFD method, system (\ref{SFD-General}) is divided into two smaller (simpler) subproblems using the framework of first-order splitting methods \cite{farago2005splitting}. The first subproblem is simply (\ref{system}) and the second subproblem is linear and models the influence of the feedback control and the low-pass time filter (\textit{i.e.} its expression is identical to (\ref{SFD-General}) without the term $\bm{\mbox{F}}(\bm{\mbox{q}})$). Hence at step $(n+1)$, the solution of the controlled system is given by
\begin{equation}
\begin{pmatrix} \bm{\mbox{q}}^{n+1} \\  \bar{\bm{\mbox{q}}}^{n+1} \end{pmatrix} 
= e^{\mathcal{L} \Delta t}
\begin{pmatrix} \bm{\Phi} (\bm{\mbox{q}}^n) \\  \bar{\bm{\mbox{q}}}^{n} \end{pmatrix},
\label{SplittingSFD}
\end{equation}
where $\Delta t$ is the time-step used within the solver $\bm{\Phi}$ and, if we note $I$ is the identity matrix, the linear operator $\mathcal{L}$ is such that
\begin{equation}
\mathcal{L} = \begin{pmatrix} - \chi I &  \chi I \\  I/\Delta & -I/\Delta
\end{pmatrix}.
\label{LinearOperator}
\end{equation}

This method is called encapsulated because the existing unsteady time-stepper $\bm{\Phi}$ does not need to be modified. This solver is simply treated as a ``black box'' and the method can be implemented as a wrapper function around it. The action of this wrapper is only to apply the linear operator $e^{\mathcal{L} \Delta t}$ to the output of $\bm{\Phi}$ and to the filtered quantity $\bar{\bm{\mbox{q}}}^{n}$ at every time step.

System (\ref{SplittingSFD}) does not converge towards the steady-state solution of (\ref{system}) for arbitrary control coefficient $\chi$ and filter width $\Delta$. These parameters have to be carefully chosen to control the evolution of the least stable mode of (\ref{system}). In Sec. \ref{SectionStabAnal} we present a method to evaluate this least stable mode. Then in Sec. \ref{SectionOptParam} we show how optimum parameters $\chi$ and $\Delta$ can be selected once the dominant eigenvalue of (\ref{system}) is known.


\section{Global stability analysis using time-stepping}
\label{SectionStabAnal}

In this section we present an overview of the main aspects of global stability problems in fluid dynamics. System (\ref{system}) models any dynamical system, but here we only focus on the incompressible Navier-Stokes equations
\begin{equation}
\begin{cases}
\partial_t \bm{\mbox{u}} = - \left( \bm{\mbox{u}} \cdot \nabla \right) \bm{\mbox{u}} - \nabla p + Re^{-1} \nabla^2 \bm{\mbox{u}}, \\
\nabla \cdot \bm{\mbox{u}} = 0,
\end{cases}
\label{IncNS}
\end{equation}
where \textbf{u} is the fluid velocity and $p$ is the modified (or kinematic) pressure. $Re = \mbox{u}_{\infty}L/\nu$ is the Reynolds number where $\mbox{u}_{\infty}$ is the mean velocity, $L$ is the characteristic length and $\nu$ is the kinematic viscosity. Appropriate initial and boundary conditions must also be defined. 

The starting point of linear stability analysis is to define a base flow \textbf{U} which is a solution of (\ref{IncNS}). This base flow can be time-dependent (\textit{e.g.} Floquet stability analysis requires a time-periodic base flow) but in this study, we are interested in steady base flows. In this section, we assume that a steady-state solution of (\ref{IncNS}) is known.

Subject to appropriate initial and boundary conditions, the evolution of infinitesimal perturbations \textbf{u}$'$ of the base flow \textbf{U} is governed by the linearised Navier-Stokes equations
\begin{equation}
\begin{cases}
\partial_t \bm{\mbox{u}}' = - \left( \bm{\mbox{U}} \cdot \nabla \right) \bm{\mbox{u}}' - \left( \bm{\mbox{u}}' \cdot \nabla \right) \bm{\mbox{U}} - \nabla p' + Re^{-1} \nabla^2 \bm{\mbox{u}}', \\
\nabla \cdot \bm{\mbox{u}}' = 0.
\end{cases}
\label{IncN_LinS}
\end{equation}

A linear evolution operator $\mathcal{A}$ can be defined from Eq. (\ref{IncN_LinS}). Hence an initial perturbation \textbf{u}$'(0)$ evolves forward in time such that
\begin{equation}
\bm{\mbox{u}}'(t) = \mathcal{A}(t) \bm{\mbox{u}}'(0).
\label{ForwPert}
\end{equation}

Note that to correctly set up system (\ref{ForwPert}), the inflow boundary conditions of the non-linear problem have to be modified to be homogeneous Dirichlet. 

We are interested in normal mode solutions of the form \textbf{u}$'$(\textbf{x}, $t$)=exp($\mu_j t$) $\hat{\bm{\mbox{u}}}_j$(\textbf{x}) + c.c., where $\hat{\bm{\mbox{u}}}_j$ are the eigenmodes and $\mu_j$ are the eigenvalues (both are, in general, complex). We can define the growth rate $\sigma_j$ and the frequency $f_j$ such that $\mu_j = \sigma_j + {\rm{i}} f_j$. 
We also define the time $T_{\mbox{\tiny{Arnoldi}}}$ as being the length of an Arnoldi iteration. $T_{\mbox{\tiny{Arnoldi}}}$ has to be large enough to allow the perturbation to evolve because large times increase the spectral gap of $\mathcal{A}(T_{\mbox{\tiny{Arnoldi}}})$, which increases the convergence rate of the method. For a given time $T_{\mbox{\tiny{Arnoldi}}}$, we obtain the eigenvalue problem
\begin{equation}
\mathcal{A}(T_{\mbox{\tiny{Arnoldi}}}) \hat{\bm{\mbox{u}}}_j = \lambda_j \hat{\bm{\mbox{u}}}_j, \mbox{~~with~~} \lambda_j = {\rm{exp}}(\mu_j T_{\mbox{\tiny{Arnoldi}}}).
\end{equation} 

Linear stability (or instability) of the base flow \textbf{U} is determined by the dominant eigenvalue of $\mathcal{A}(T_{\mbox{\tiny{Arnoldi}}})$ (\textit{i.e.} the one of largest modulus). If there exists at least one $\lambda_j$ such that $|\lambda_j| > 1$, then infinitesimal perturbations of the base flow exponentially grow in time. Hence \textbf{U} is said to be asymptotically linearly unstable. In opposition, if all the eigenvalues verify $|\lambda_j| < 1$, then infinitesimal perturbations decay in time and the base flow is linearly stable. Most of the time $|\lambda_j| = 1$ indicates a bifurcation point. Note that instead of evaluating the modulus of the eigenvalues $\lambda_j$, one can also look at the sign of the corresponding growth rate $\sigma_j$. A negative growth rate corresponds to a linearly stable base flow and a positive growth rate, to an linearly unstable one.

To compute the eigenvalues of $\mathcal{A}(T_{\mbox{\tiny{Arnoldi}}})$, the ``time-stepper'' approach \cite{Bifurcation_Timesteppers, ModifiedArnoldi} is used. The idea is to perform stability analysis by adapting an existing Navier-Stokes code that solves (\ref{IncNS}). Note that in Sec. \ref{SectionEncapsSFD} a comparable strategy was used, where the encapsulated SFD method was presented as being easy to implement as a wrapper around an existing unsteady solver.

Dominant eigenvalues are obtained using the modified Arnoldi iteration method introduced by Barkley \textit{et al} \cite{ModifiedArnoldi}. 
The mechanism of this algorithm is that the operator $\mathcal{A}(T_{\mbox{\tiny{Arnoldi}}})$ is applied several times to a random initial vector \textbf{u}$_0$ (which is non-zero). This generates a Krylov subspace and then an upper Hessenberg matrix \textbf{H}. The dominant eigenvalue of \textbf{H} converges towards the dominant eigenvalue of $\mathcal{A}(T_{\mbox{\tiny{Arnoldi}}})$ in a relatively limited number of iterations. The computational cost and memory requirements of this algorithm are low. The most demanding part is to compute $\mathcal{A}(T_{\mbox{\tiny{Arnoldi}}})$ every Arnoldi iteration. 

This algorithm has been recently implemented into the Nektar++ spectral/\textit{hp} element framework \cite{Nektar++}. Rocco \cite{GabrieleThesis} showed that the results agree very well with ARPACK (ARnoldi PACKage \cite{Arpack}), the differences between the dominant eigenvalues obtained with the two methods were found to be smaller that $10^{-5}$. 

Note that, when an adjoint linearised system is implemented, this algorithm can be adapted to compute transient growth analysis and evaluate the influence of non-normal modes.


\section{Evaluation of optimum parameters}
\label{SectionOptParam}

Global stability analysis aims to determine if the largest eigenvalue of the linear operator $\mathcal{A}(T_{\mbox{\tiny{Arnoldi}}})$ has a modulus larger than one or not. When the SFD method is applied to a system, it aims to control the evolution of the least stable eigenvalue of this system. Hence determining if the SFD method can control a dynamical system may be narrowed down to: can the SFD method control the evolution of the dominant instability of this system?

At this stage, we assume the dominant eigenvalue (denoted $\lambda_D$) of the unstable dynamical system studied is known. We introduce the one-dimensional model  
\begin{equation}
u^{n+1} = \lambda_D u^n.
\label{1DModel}
\end{equation}

For a given control coefficient $\chi$ and filter width $\Delta$, if the SFD method is able to force (\ref{1DModel}) to evolve towards its steady-state (which is $u=\bar{u}=0$), it means that it is able to control the instabilities related to $\lambda_D$. Then the same parameters can be used in the SFD method applied to an unstable fluid system which has $\lambda_D$ for dominant eigenvalue. The encapsulated formulation of the SFD method is applied to (\ref{1DModel}), is such that
\begin{equation}
\begin{pmatrix}
u^{n+1} \\  \bar{u}^{n+1}
\end{pmatrix} = e^{\mathcal{L}_{1\text{D}}}
\begin{pmatrix} \lambda_D u^n \\  \bar{u}^{n} \end{pmatrix}
= \underbrace{e^{\mathcal{L}_{1\text{D}}}
\begin{pmatrix} \lambda_D & 0\\  0 & 1
\end{pmatrix}}_{\mathcal{M}}
\begin{pmatrix} u^n \\  \bar{u}^{n} \end{pmatrix},
\label{1D_SplittingSFD}
\end{equation}
where $\mathcal{L}_{1\text{D}}$ is similar to (\ref{LinearOperator}), with $I$ replaced by $1$, and  $\mathcal{M}$ is the iteration matrix transforming $(u^n,\bar{u}^n)$ into $(u^{n+1},\bar{u}^{n+1})$. $\mathcal{M}$ depends on $\lambda_D$, $\chi$ and $\Delta$.

The convergence of (\ref{1D_SplittingSFD}) is only governed by the two eigenvalues of $\mathcal{M}$. If they both have a modulus strictly smaller than one, then system (\ref{1D_SplittingSFD}) is stable and converges towards its steady-state solution. Otherwise, the system is unstable and the steady-state can not be reached. 

As $\mathcal{M}$ is a 2 by 2 matrix, evaluating the modulus of its eigenvalues is simple. For a fixed $\lambda_D$, we can manually select a control coefficient $\chi$ and a filter width $\Delta$ that ensure both eigenvalues of $\mathcal{M}$ to be smaller than one. Then the instabilities are controlled by the SFD method and convergence of (\ref{1D_SplittingSFD}) towards its steady-state is guaranteed. However, this convergence may be slow.

The fastest convergence of (\ref{1D_SplittingSFD}) is achieved when the modulus of the dominant eigenvalue of $\mathcal{M}$ is minimum. As we can evaluate the eigenvalues of $\mathcal{M}$ for every $\chi \geqslant 0$ and every $\Delta > 0$ (for a fixed $\lambda_D$), a basic line search algorithm can be implemented to obtain the optimum parameters $\chi_{\rm{opt}}$ and $\Delta_{\rm{opt}}$ that give the minimum of the modulus of the dominant eigenvalue of $\mathcal{M}$. Then these parameters can be used in the SFD method applied to a flow which has $\lambda_D$ for dominant eigenvalue.

The issue with this feature is that to be able to adjust the parameters of the SFD method, the dominant eigenvalue $\lambda_D$ of the system is required. But this eigenvalue is obtained only after computing stability analysis; and stability analysis requires to linearise the governing equation around a steady-state solution. Which is what we intend to reach with the SFD method. In other words, to ensure convergence of the SFD method towards a steady-state solution, we need to know this steady-state solution. Fig. \ref{ClosedLoop} illustrates this implicit problem. The closed loop shows that analysing the stability of the one-dimensional model ($\ref{1DModel}$) is useful only if the dominant eigenvalue $\lambda_D$ is known \textit{a priori}.
\tikzstyle{decision} = [diamond, draw, fill=green!8, text width=5.0em, text badly centered, inner sep=0pt]
\tikzstyle{block} = [rectangle, draw, fill=blue!8, text width=5.5em, text centered, rounded corners, minimum height=3.25em]
\tikzstyle{cloud} = [draw, ellipse,fill=red!8, minimum height=2em]
\tikzstyle{point} = [circle, draw, fill=white!100, minimum height=0em]
\tikzstyle{line} = [draw, -latex'] \tikzstyle{cloud} = [draw, ellipse,fill=red!20, node distance=3cm, minimum height=2em]
\begin{figure}[h]
\begin{center}
\begin{tikzpicture}[node distance = 2.75cm, auto]
    \node [block] (EV) {$\lambda_D$ known};
    \node [block, right of=EV] (1D) {1D Model};
    \node [block, right of=1D] (Opt) {Obtain $\chi_{\rm{opt}}$ and $\Delta_{\rm{opt}}$};
    \node [block, below of=Opt, node distance=2.5cm] (SFD) {Execute SFD};
    \node [block, left of=SFD] (SS) {Steady-state Solution};
    \node [block, left of=SS] (StabAnalysis) {Stability Analysis};
    \path [line] (EV) -- (1D);
    \path [line] (1D) -- (Opt);
    \path [line] (Opt) -- (SFD);
    \path [line] (SFD) -- (SS);
    \path [line] (SS) -- (StabAnalysis);
    \path [line] (StabAnalysis) -- (EV);
\end{tikzpicture}
\caption{\label{ClosedLoop} Illustration of the implicit problem of the unadapted SFD method. $\lambda_D$ is the dominant eigenvalue of the flow; $\chi_{\rm{opt}}$ and $\Delta_{\rm{opt}}$ are the parameters that ensure an optimum convergence of the SFD method applied to ($\ref{1DModel}$).}
\end{center}
\end{figure}
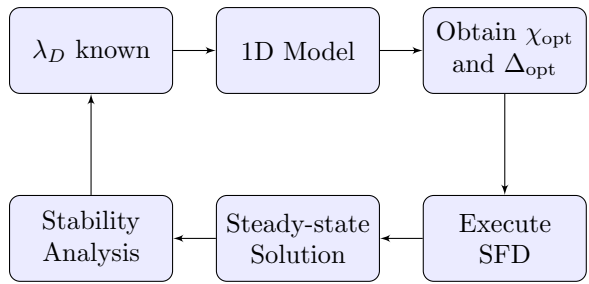

In the following section we present an adaptive algorithm to get around this issue. 


\section{Adaptive algorithm}
\label{SectionAdaptive}

In this section, we propose a procedure to address the issue of selecting appropriate parameters for the SFD method. Our algorithm links together the SFD method presented in Sec \ref{SectionEncapsSFD}, the stability analysis method summarised in Sec. \ref{SectionStabAnal} and the one-dimensional model introduced in Sec. \ref{SectionOptParam}. The idea is to use a ``partially converged'' steady base flow for computing stability analysis and obtain an approximation of the dominant eigenvalue (denoted $\tilde{\lambda}$). Then we use the procedure described in Sec. \ref{SectionOptParam} to obtain the control coefficient $\tilde{\chi}$ and the filter width $\tilde{\Delta}$ that ensure the fastest convergence of the SFD method applied to $u^{n+1} = \tilde{\lambda} u^n$. These parameters are an approximation of the optimum parameters $\chi_{\rm{opt}}$ and $\Delta_{\rm{opt}}$. The SFD method is then executed using the parameters $\tilde{\chi}$ and $\tilde{\Delta}$, so the instabilities can be damped more efficiently.

The main aspect of the procedure proposed here is the selection of the ``partially converged'' steady-state. In this study the choice made was to execute the SFD method for a given number $T$ of time units before recomputing the control parameters. The user has to define $T$ considering that it has to be large enough to allow the flow to evolve but not too large to avoid wasting computational time executing the SFD method with badly suited parameters.

After defining the ``partially converged'' steady-state, the stability analysis method has to be computed. Note that here, the base flow used is not exactly a steady-state solution of the governing equations. The role of the SFD method is to reduce the temporal frequencies within the flow. Hence as the problem is not converged yet, the base flow selected is only an approximation of the steady-state solution. However our experiments suggested that this analysis gives a good approximation of the dominant eigenvalue of the flow studied. The user has to define the tolerance $\varepsilon_{\mbox{\tiny{Stab}}}$ to determine when the modified Arnoldi iteration method is converged. As we only seek an approximation of the dominant eigenvalue, this parameter does not need to be very small. 

At this stage, the one-dimensional model presented in Sec. \ref{SectionOptParam} is used to approximate the optimum parameters. After updating the parameters $\tilde{\chi}$ and $\tilde{\Delta}$, the SFD method is executed again for $T$ time units. 

This adaptive process is iterated until the norm $||\bm{\mbox{q}} - \bar{\bm{\mbox{q}} }||_{\rm{inf}}$ becomes smaller than a desired tolerance $\varepsilon_{\mbox{\tiny{Adapt}}}$. When this becomes true, we consider that the approximation of the steady base flow is good enough, hence $\tilde{\chi}$ and $\tilde{\Delta}$ are fixed until convergence is reached. The tolerance $\varepsilon_{\mbox{\tiny{Adapt}}}$ has to be defined by the user too. If this tolerance is too large, then the flow will not have enough time to evolve, and the ``partially converged'' flow will not be a good approximation of the steady-state. If this tolerance is too small, stability analysis will be computed a lot of times using base flows close to each other which will waste computational resources. 

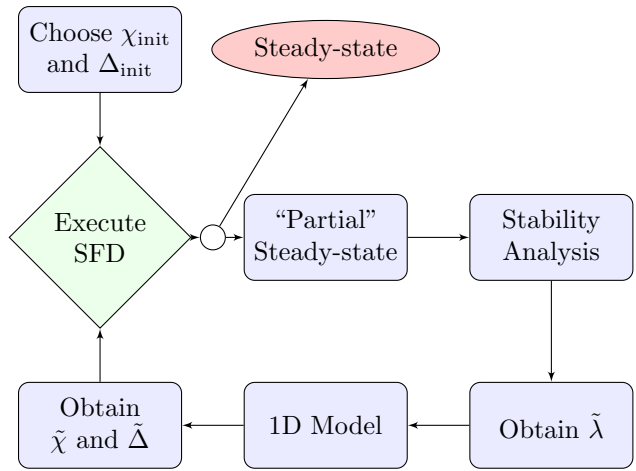
\begin{figure}[h]
\begin{center}
\begin{tikzpicture}[node distance = 3.0cm, auto]
    \node [block] (Random) {Choose $\chi_{\rm{init}}$ and $\Delta_{\rm{init}}$};
    \node [decision, below of=Random, node distance = 2.5cm] (SFD) {Execute SFD};
    \node [point, right of=SFD, node distance = 1.5cm] (Check) {};
    \node [block, right of=Check, node distance = 1.5cm] (Part) {``Partial'' Steady-state};
    \node [cloud, above of=Part, node distance = 2.5cm] (SS) {Steady-state};
    \node [block, right of=Part] (StabAnalysis) {Stability Analysis};
    \node [block, below of=StabAnalysis, node distance = 4.75cm, node distance = 2.5cm] (ApproxEV) {Obtain $\tilde{\lambda}$};
    \node [block, left of=ApproxEV] (1D) {1D Model};
    \node [block, left of=1D] (ApproxParam) {Obtain $\tilde{\chi}$ and $\tilde{\Delta}$};
    \path [line] (Random) -- (SFD);
    \path [line] (SFD) -- (Check);
    \path [line] (Check) -- (Part);
    \path [line] (Part) -- (StabAnalysis);
    \path [line] (StabAnalysis) -- (ApproxEV);
    \path [line] (ApproxEV) -- (1D);
    \path [line] (1D) -- (ApproxParam);
    \path [line] (ApproxParam) -- (SFD);
    \path [line] (Check) -- (SS);
\end{tikzpicture}
\caption{\label{OpenLoop} Illustration of the adaptive SFD method. $\chi_{\rm{init}}$ and $\Delta_{\rm{init}}$  are the initial parameters of the SFD method; $\tilde{\lambda}$ is an approximation of the dominant eigenvalue of the flow; $\tilde{\chi}$ and $\tilde{\Delta}$ are the parameters that ensure an optimum convergence of the SFD method applied to $u^{n+1} = \tilde{\lambda} u^n$. Note that the circle (right of ``Execute SFD'') represents a group of criteria that determine if the stability analysis method has to be computed or if the SFD parameters are fixed until convergence is reached.}
\end{center}
\end{figure}

This adaptive method is illustrated on Fig. \ref{OpenLoop}. Note that when stability analysis has to be computed more than once (\textit{i.e.} the loop of Fig. \ref{OpenLoop} is executed several times), the initial condition of the modified Arnoldi iteration method is the final solution of the previous computation. This allows us to speed up convergence. 

Also, we recall that the SFD method can only control unstable problems with oscillatory growth of the instabilities (\textit{i.e.} the imaginary part of the dominant eigenvalue is non-zero). Hence all the $\tilde{\lambda}$ successively computed for this algorithm must have a non-zero imaginary part as well.

All users of the SFD method (in any formulation) must define an initial control coefficient and filter width. To execute the adaptive SFD method, one extra parameter (the time $T$ between two consecutive execution of the stability analysis method) has to be defined. This time interval plays a key role in the ability of the method to quickly approach the optimum parameters $\chi_{\rm{opt}}$ and $\Delta_{\rm{opt}}$ in order to reach the fastest convergence rate. Two tolerances $\varepsilon_{\mbox{\tiny{Stab}}}$ and $\varepsilon_{\mbox{\tiny{Adapt}}}$, which respectively define the stability analysis method residual tolerance and the limit such that the SFD parameters are fixed when $||\bm{\mbox{q}} - \bar{\bm{\mbox{q}} }||_{\rm{inf}}<\varepsilon_{\mbox{\tiny{Adapt}}}$, must also be set up. 



\section{Numerical simulations}
\label{SectionNumSim}

In this section we present the application of the adaptive SFD method detailed in Sec. \ref{SectionAdaptive} to some classical examples of computational fluid dynamics. The aim is to show that the method is able to reach unstable steady-state solutions without requiring any \textit{a priori} knowledge of the flow stability behaviour.

The algorithm has been implemented into the Nektar++ spectral/\textit{hp} element framework \cite{Nektar++}. It couples two methods that were already present into this package which are the encapsulated SFD method and a modified Arnoldi iteration method. The incompressible Navier-Stokes solver used throughout this section implements the \textit{velocity-correction} scheme with an unstabilised continuous Galerkin method\cite{SencerBook} to discretize the problem in space and a second order implicit-explicit (IMEX) method \cite{ascher1995implicit} for time-integration scheme.

In our implementation, the user has to define the initial parameters $\chi_{\rm{init}}$ and $\Delta_{\rm{init}}$, the time $T$ for which the SFD method is executed before computing stability analysis, the tolerance $\varepsilon_{\mbox{\tiny{Stab}}}$ of the modified Arnoldi iteration method and the tolerance $\varepsilon_{\mbox{\tiny{Adapt}}}$ which determines when we stop adapting the control coefficient and the filter width. Throughout this section, we fix $\varepsilon_{\mbox{\tiny{Stab}}} = 10^{-3}$, $\varepsilon_{\mbox{\tiny{Adapt}}} = 10^{-2}$ and we considered the problem to be converged when $||\bm{\mbox{q}} - \bar{\bm{\mbox{q}} }||_{\rm{inf}}<10^{-8}$.

Also, we define the length of one Arnoldi iteration to be equal to one time unit. Hence if the stability analysis converges in $n$ iterations, the computational time will be approximately the same as executing the Navier-Stokes solver for $n$ iterations.

Four two-dimensional simulations are presented in this section. In Sec. \ref{SubSecRe100} the behaviour of the adaptive SFD method is detailed for the incompressible flow past a cylinder at $Re=100$. Then in Sec. \ref{SubSecRe300} we show that this case can easily be extended to higher Reynolds numbers and present the case $Re=300$. In Sec. \ref{SectionEllipse}, the incompressible flow past an ellipse at $Re=150$ is also studied in order to show that the adaptive SFD method works well on geometries that do not contain axial symmetries. Finally in Sec. \ref{SectionRotatingCyl} we consider the incompressible flow past a rotating cylinder at $Re=100$ and a rotation rate $\alpha = 5$, and show that the steady-state solution of the unstable mode II can be reached without requiring the use of continuation methods.


\subsection{Incompressible flow past a cylinder at $Re=100$}
\label{SubSecRe100}

The first test case presented is the two-dimensional incompressible flow past a circular cylinder at $Re = 100$ (where the diameter of the cylinder is the characteristic length). At this Reynolds number, the flow is unstable and, if no control method is used, vortex shedding occurs. It has been shown \cite{Encaps_SFD} that the SFD method is able to suppress the unsteady oscillations in the cylinder wake and force the system to evolve towards its steady-state solution. In this section we present the execution of the adaptive algorithm presented in Sec. \ref{SectionAdaptive} for two different pairs of initial parameters: one for which the unadapted SFD method is converging slowly and the other for which the unadapted SFD method is not converging. Our goal is to show that, for both cases, with the adaptive procedure we can approach the optimum convergence rate of the SFD method.

Both cases presented in this section use the same computational domain. It is composed of 746 elements and its dimensions are $-15 \leq x \leq 45$ and $-25 \leq y \leq 25$. The mesh is made of structured curved quadrilaterals close to the cylinder boundary and triangles elsewhere. No-slip boundary conditions are imposed at the cylinder surface and Dirichlet boundary conditions $(u,~v) = (1,~0)$ are set at the left, top and bottom edges. An outflow boundary condition is defined at the right edge of the domain. The initial conditions are such that $(u_0,~v_0) = (0,~0)$. As the solution is expected to be smooth, a polynomial order of 7 is used and the time-step is $\Delta t = 0.01$. 



For the first test case, we execute the adaptive SFD method with initial parameters that ensure a slow convergence of the unadapted SFD method. The initial parameters of the SFD method are $\chi_{\rm{init}} = 1$ and $\Delta_{\rm{init}} = 1$. With these values, the unadapted method converges very slowly (in about 4100 time units). Then here, we aim to improve the convergence rate of the SFD method by updating the control coefficient and the filter width during the solver execution. 

First we choose $T=25$ to execute our adaptive algorithm, which means that the stability analysis method will be computed every 25 time units until $||\bm{\mbox{q}} - \bar{\bm{\mbox{q}} }||_{\rm{inf}}$ becomes smaller than $\varepsilon_{\mbox{\tiny{Adapt}}}=10^{-2}$. In this case, stability analysis is executed 6 times for a total of 225 Arnoldi steps. We recall that for the modified Arnoldi iteration method, the computational cost of one step is approximately the same as when the non-linear solver executed. The steady-state solution is reached after a total of 1093.5 time units (including the Arnoldi steps). Hence about $20\%$ of the computational resources are spent for evaluating the dominant eigenvalue at several ``partially converged'' base flows. This is eventually a good investment because the total number of steps required to converge is nearly four times smaller than when the unadapted SFD method is executed. 
\begin{table*}
{\small{
\begin{tabular}{|c|c|c|c|c|}
\hline 
\rule[-1ex]{0pt}{2.5ex} Time & Arnoldi method & Dominant eigenvalue & SFD parameters & ~ $||\bm{\mbox{q}} - \bar{\bm{\mbox{q}} }||_{\rm{inf}}$ ~ \\ 
\hline 
\rule[-1ex]{0pt}{2.5ex} $t=25$ & ~ Conv. in 59 steps ~ & ~ $\tilde{\sigma}_1 = 0.135$; $\tilde{f}_1 = 0.908$ ~ & ~ $\tilde{\chi}_1 = 0.548$; $\tilde{\Delta}_1 = 2.482$ ~ & 0.0399 \\ 
\hline 
\rule[-1ex]{0pt}{2.5ex} $t=109$ & Conv. in 39 steps & $\tilde{\sigma}_2 = 0.143$; $\tilde{f}_2 = 0.823$ & $\tilde{\chi}_2 = 0.506$; $\tilde{\Delta}_2 = 2.821$ & 0.0355 \\ 
\hline 
\rule[-1ex]{0pt}{2.5ex} $t=173$ & Conv. in 38 steps & $\tilde{\sigma}_3 = 0.136$; $\tilde{f}_3 = 0.785$ & $\tilde{\chi}_3 = 0.481$; $\tilde{\Delta}_3 = 2.967$ & 0.0180 \\ 
\hline 
\rule[-1ex]{0pt}{2.5ex} $t=236$ & Conv. in 34 steps & $\tilde{\sigma}_4 = 0.133$; $\tilde{f}_4 = 0.766$ & $\tilde{\chi}_4 = 0.468$; $\tilde{\Delta}_4 = 3.042$ & 0.0158 \\ 
\hline 
\rule[-1ex]{0pt}{2.5ex} $t=295$ & Conv. in 33 steps & $\tilde{\sigma}_5 = 0.130$; $\tilde{f}_5 = 0.755$ & $\tilde{\chi}_5 = 0.461$; $\tilde{\Delta}_5 = 3.084$ & 0.0122 \\ 
\hline 
\rule[-1ex]{0pt}{2.5ex} ~ $t=335$ ~ & Conv. in 22 steps & $\tilde{\sigma}_6 = 0.129$; $\tilde{f}_6 = 0.753$ & $\tilde{\chi}_6 = 0.459$; $\tilde{\Delta}_6 = 3.193$ & 0.0099 \\ 
\hline 
\end{tabular} 
}}
\caption{\label{TableT_25}Successive execution of the stability analysis method for the adaptive SFD method when $T=25$ and $\chi_{\rm{init}} = 1$ and $\Delta_{\rm{init}} = 1$. The time at which the modified Arnoldi iteration method is called and the number of steps required to converge (\textit{i.e.} when the residual becomes smaller than $10^{-3}$) are reported in the first and second columns. The third column lists the successive approximations of the dominant eigenvalue, the corresponding optimum parameters are reported in the fourth column and the norm $||\bm{\mbox{q}} - \bar{\bm{\mbox{q}} }||_{\rm{inf}}$ at each execution of the modified Arnoldi iteration method in the fifth column.}
\end{table*}

\begin{figure}[h]
\begin{center}
\includegraphics[width=8cm]{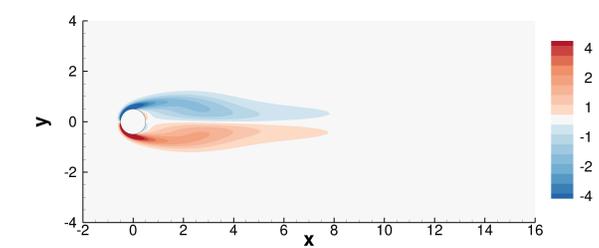}
\caption{\label{BF1_Conv} Vorticity of the first ``partially converged'' steady-state. Obtained for the parameters $\chi_{\rm{init}} = 1$, $\Delta_{\rm{init}} = 1$ and $T=25$.}
\end{center}
\end{figure}

To illustrate the algorithm behaviour in more detail, the results of the successive execution of the stability analysis method are reported in Table \ref{TableT_25}. The first ``partially converged'' base flow (obtained at $t=25$) is shown on Fig. \ref{BF1_Conv}. The first line of Table \ref{TableT_25} is related to this flow. 

\begin{figure}[h]
\begin{center}
\includegraphics[width=8cm]{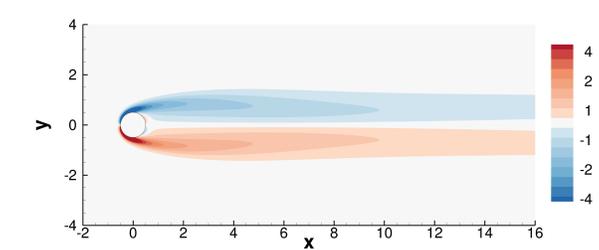}
\caption{\label{Fig_Re100_SS} Vorticity of the  steady-state of the incompressible flow past a two-dimensional cylinder at $Re=100$.}
\end{center}
\end{figure}

In Table \ref{TableT_25}, we observe that after the third iteration, the value of the approximated eigenvalue and the corresponding SFD parameters evolve very little. This suggests that, for this problem, a coarse approximation of the steady-state solution provides a quite good evaluation of the dominant eigenvalue of the flow. When convergence is reached (\textit{i.e.} $||\bm{\mbox{q}} - \bar{\bm{\mbox{q}} }||_{\rm{inf}}<10^{-8}$), the steady-state solution obtained is shown on Fig. \ref{Fig_Re100_SS} and corresponds to Barkley's result \cite{BarkleyMeanVsSteadyCylinderFlow}.

We computed an \textit{a posteriori} stability analysis using the converged steady-state solution as base flow. The growth rate and frequency of this flow were found to be $\sigma = 0.127$ and $f = 0.741$ \footnote{Note that here, a second order IMEX scheme was used to execute the modified Arnoldi iteration method. Jordi \textit{et al.}\cite{Encaps_SFD} used a first order IMEX scheme to obtain this eigenvalue, which explains why they are slightly different.}. Knowing this dominant eigenvalue, we can evaluate the optimum parameters to obtain the optimum convergence rate of the unadapted SFD method. The optimum control coefficient is $\chi_{\rm{opt}} = 0.451$ and the optimum filter width is $\Delta_{\rm{opt}} = 3.144$, very close to the parameters presented on the last line of Table \ref{TableT_25}. If the unadapted SFD method is executed using these optimum parameters, it converges in 878 time units, which is only $20\%$ faster than using the adaptive algorithm. Hence the adaptive algorithm used here can approach the optimum convergence rate of the SFD method without requiring any \textit{a priori} knowledge of the flow.

We executed other simulations only changing the time $T$ between two consecutive executions of the modified Arnoldi iteration method. For $T=50$, convergence is reached in 996 time units (including 134 Arnoldi steps) and details of the three successive executions of the stability analysis method are reported in Table \ref{TableT_50}. We also observe for this case that the approximations of the dominant eigenvalue are rapidly accurate and that the final SFD parameters are very close to the optimum ones. 
\begin{table*}
{\small{
\begin{tabular}{|c|c|c|c|c|}
\hline 
\rule[-1ex]{0pt}{2.5ex} Time & Arnoldi method & Dominant eigenvalue & SFD parameters & ~ $||\bm{\mbox{q}} - \bar{\bm{\mbox{q}} }||_{\rm{inf}}$ ~ \\ 
\hline 
\rule[-1ex]{0pt}{2.5ex} $t=50$ & ~ Conv. in 59 steps ~ & ~ $\tilde{\sigma}_1 = 0.142$; $\tilde{f}_1 = 0.813$ ~ & ~ $\tilde{\chi}_1 = 0.499$; $\tilde{\Delta}_1 = 2.859$ ~ & 0.0225 \\ 
\hline 
\rule[-1ex]{0pt}{2.5ex} $t=159$ & Conv. in 41 steps & $\tilde{\sigma}_2 = 0.132$; $\tilde{f}_2 = 0.763$ & $\tilde{\chi}_2 = 0.466$; $\tilde{\Delta}_2 = 3.054$ & 0.0158 \\ 
\hline 
\rule[-1ex]{0pt}{2.5ex} ~ $t=228$ ~ & Conv. in 34 steps & $\tilde{\sigma}_3 = 0.130$; $\tilde{f}_3 = 0.753$ & $\tilde{\chi}_3 = 0.459$; $\tilde{\Delta}_3 = 3.095$ & 0.0099 \\ 
\hline 
\end{tabular} 
}}
\caption{\label{TableT_50}Successive execution of the stability analysis method for the adaptive SFD method when $T=50$ and $\chi_{\rm{init}} = 1$ and $\Delta_{\rm{init}} = 1$. The time at which the modified Arnoldi iteration method is called and the number of steps required to converge (\textit{i.e.} when the residual becomes smaller than $10^{-3}$) are reported in the first and second columns. The third column lists the successive approximations of the dominant eigenvalue, the corresponding optimum parameters are reported in the fourth column and the norm $||\bm{\mbox{q}} - \bar{\bm{\mbox{q}} }||_{\rm{inf}}$ at each execution of the modified Arnoldi iteration method in the fifth column.}
\end{table*}

Note that for $\chi_{\rm{init}} = 1$ and $\Delta_{\rm{init}} = 1$, the norm $||\bm{\mbox{q}} - \bar{\bm{\mbox{q}} }||_{\rm{inf}}$ becomes smaller than $10^{-2}$ at time $t=75.5$. Hence, for $T=100$, the stability analysis method is only computed once and convergence is reached in a total of 920 time units (including 65 Arnoldi steps), which is less that $5\%$ slower than when the optimum parameters are chosen to execute the unadapted SFD method.

\begin{figure*}
\begin{center}
\subfigure[]{\label{ConvHist_Conv_Large} \includegraphics[width=8.8cm]{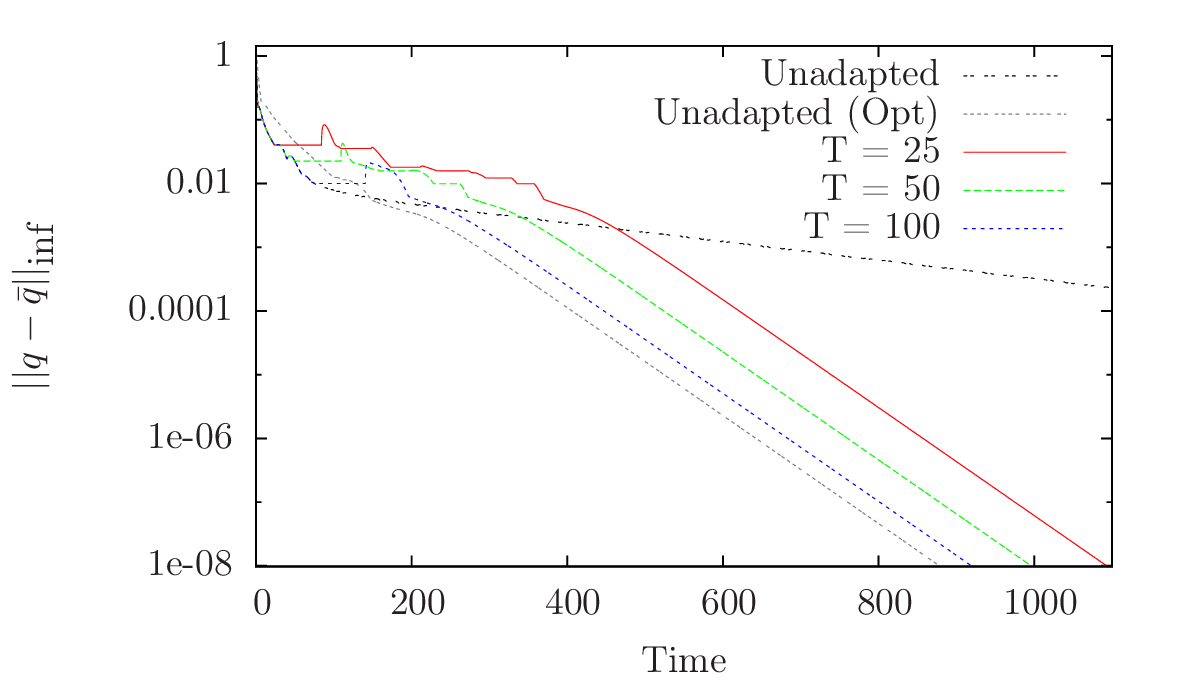}}
\subfigure[]{\label{ConvHist_Conv_Zoom} \includegraphics[width=8.8cm]{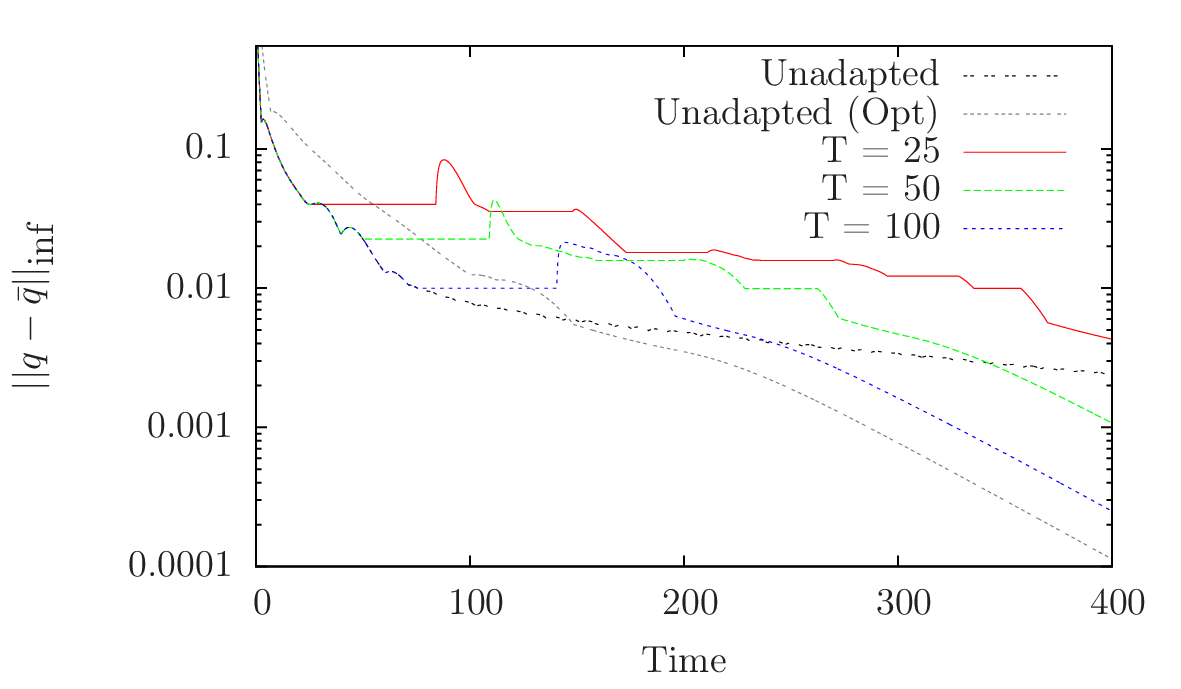}}
\caption{\label{ConvHist_Conv} Convergence history of the adapted SFD method for $T=25$, $T=50$ and $T=100$ with $\chi_{\rm{init}} = 1$ and $\Delta_{\rm{init}} = 1$. The cases of the unadapted SFD method for $\chi=1$, $\Delta=1$ (curve called ``Unadapted'') and for the optimum parameters $\chi_{\rm{opt}} = 0.4512$, $\Delta_{\rm{opt}} = 3.144$ (curve called ``Unadapted (Opt)'') are also reported.}
\end{center}
\end{figure*}

The convergence history of the cases where $T=25$, $T=50$ and $T=100$ are reported on Fig. \ref{ConvHist_Conv} along with the cases of the unadapted SFD method for $\chi=1$ and $\Delta=1$ and for the optimum parameters $\chi_{\rm{opt}} = 0.4512$ and $\Delta_{\rm{opt}} = 3.144$. The horizontal plateaux illustrate the time spent executing the stability analysis method. It is noticeable that in our adaptive method, after the control coefficient and filter width are fixed, all the cases have the same exponential convergence rate. This rate is greater than the unadapted SFD method for $\chi=1$ and $\Delta=1$ but similar to those of the unadapted SFD method executed with optimum parameters.



We now execute exactly the same set of simulations, only changing the initial parameters. We select $\chi_{\rm{init}} = 1$ and $\Delta_{\rm{init}} = 0.5$. These initial parameters do not enable the SFD method to converge \cite{Encaps_SFD}. The time evolution of $||\bm{\mbox{q}} - \bar{\bm{\mbox{q}} }||_{\rm{inf}}$ is such that it decreases for a certain time, then increases abruptly and eventually oscillates around a fixed value. Hence in this case, we aim to use the adaptive algorithm presented in Sec. \ref{SectionAdaptive} to adjust the parameters such that a steady-state solution can be reached.

The algorithm behaviour is comparable to the case where initial parameters enable convergence of the unadapted SFD method. Indeed, for $T=25$ the stability analysis method is called 6 times for a total of 206 Arnoldi steps. At each execution, the stability analysis method converge in fewer steps and the approximation of the dominant eigenvalue is of better quality (and so is the approximation of the optimum parameters). For conciseness, a detailed table is not reported in this section. 

\begin{figure}[h]
\begin{center}
\includegraphics[width=8cm]{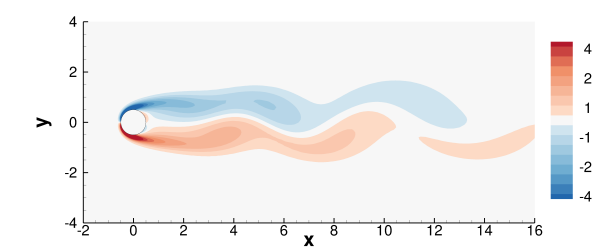}
\caption{\label{BF1_UnConv} Vorticity of the first ``partially converged'' steady-state. Obtained for the parameters $\chi_{\rm{init}} = 1$, $\Delta_{\rm{init}} = 0.5$ and $T=50$.}
\end{center}
\end{figure}

The first ``partially converged'' steady flow of the case $T=50$ is shown on Fig. \ref{BF1_UnConv}. Note that this flow configuration is not symmetric but the oscillation are reduced in comparison with the vortex shedding of the uncontrolled case. However the stability analysis executed using this ``partially converged'' base flow approximates the dominant eigenvalue of the flow well. This is an other argument to support the idea that a coarse approximation of the steady flow can lead to a good approximation of the dominant eigenvalue.

\begin{figure*}
\begin{center}
\subfigure[]{\label{ConvHist_Conv_Large} \includegraphics[width=8.8cm]{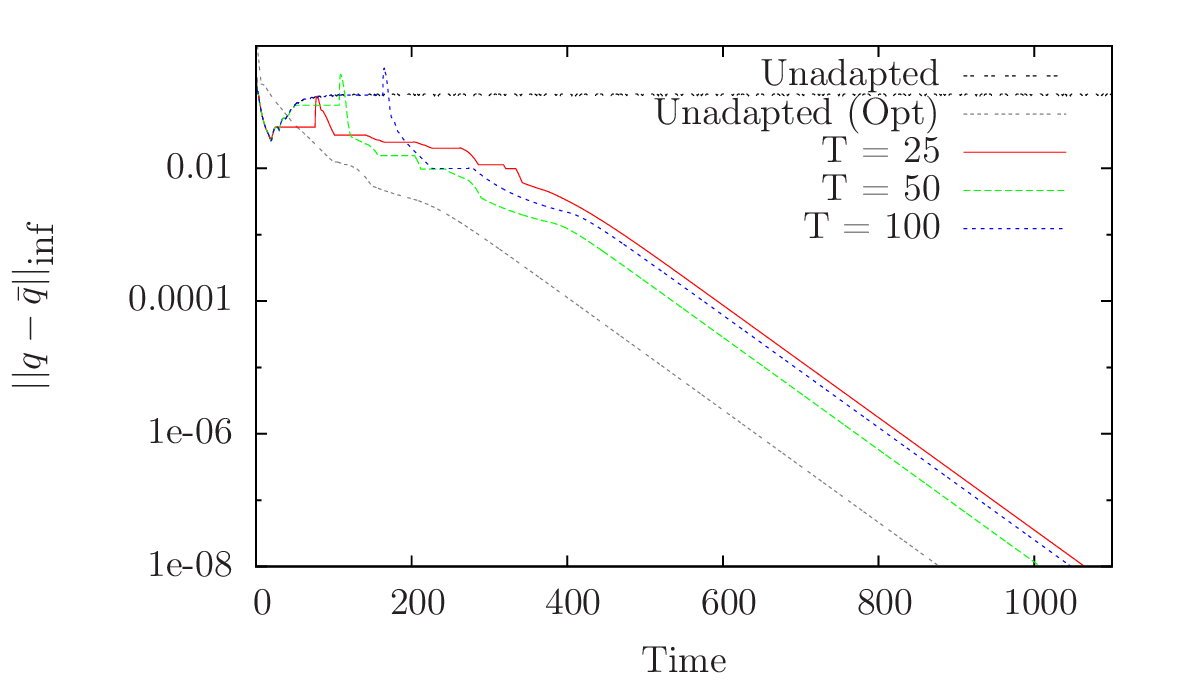}}
\subfigure[]{\label{ConvHist_Conv_Zoom} \includegraphics[width=8.8cm]{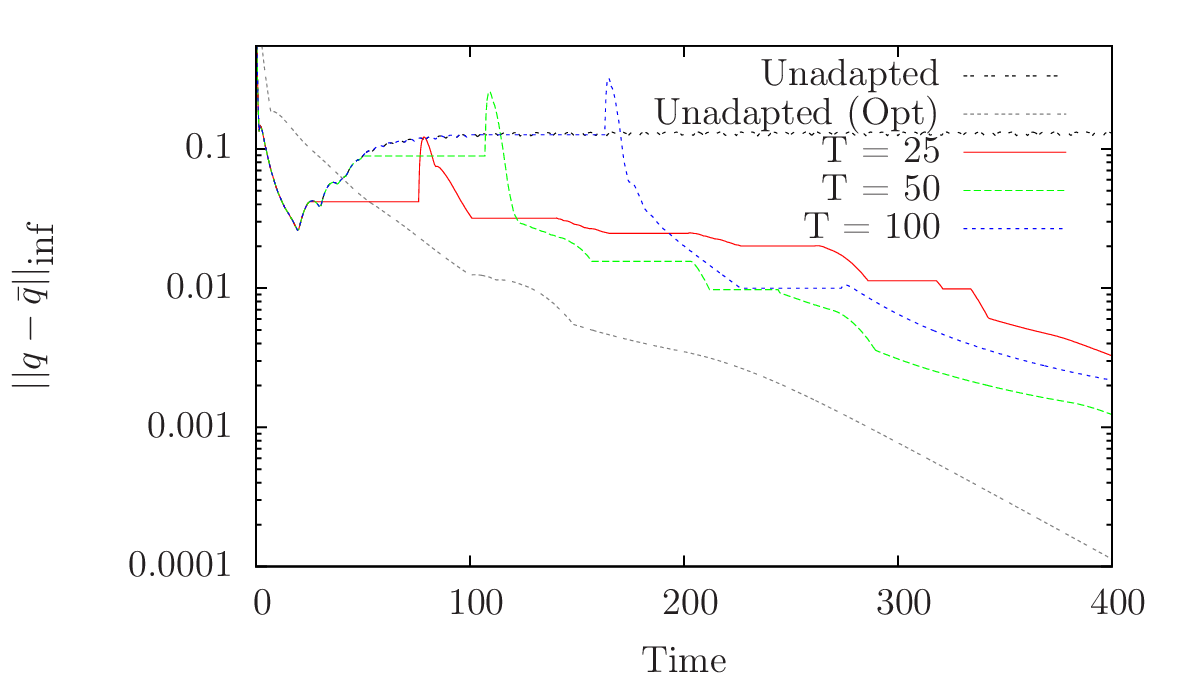}}
\caption{\label{ConvHist_Div} Convergence history of the adapted SFD method for $T=25$, $T=50$ and $T=100$ with $\chi_{\rm{init}} = 1$ and $\Delta_{\rm{init}} = 0.5$. The cases of the unadapted SFD method for $\chi=1$, $\Delta=0.5$ (curve called ``Unadapted'') and for the optimum parameters $\chi_{\rm{opt}} = 0.4512$, $\Delta_{\rm{opt}} = 3.144$ (curve called ``Unadapted (Opt)'') are also reported.}
\end{center}
\end{figure*}

The convergence history of the cases where $T=25$, $T=50$ and $T=100$ are reported on Fig. \ref{ConvHist_Div} along with the cases of the unadapted SFD method for $\chi=1$ and $\Delta=0.5$ and for the optimum parameters $\chi_{\rm{opt}} = 0.4512$ and $\Delta_{\rm{opt}} = 3.144$. The horizontal plateaux illustrate the time spent executing the stability analysis method. It is noticeable that in our adaptive method, after the control coefficient and filter width are fixed, all the cases have the same exponential convergence rate. This rate is similar to those of the unadapted SFD method executed with the optimum parameters.

We conclude this section by observing that the adaptive SFD method has successfully (and automatically) selected parameters that ensure an optimum convergence rate towards a steady-state solution even if little care is taken when choosing the initial control coefficient and filter width. 

\subsection{Incompressible flow past a cylinder at $Re=300$}
\label{SubSecRe300}


In this section we show that for the incompressible flow past a cylinder, the extension to higher Reynolds number test cases is straightforward. We execute the adaptive algorithm presented in Sec. \ref{SectionAdaptive} for $Re=300$. 

The computational domain is composed of 1330 elements and its dimensions are $-15 \leq x \leq 100$ and $-30 \leq y \leq 30$. 
The mesh is made of structured curved quadrilaterals close to the cylinder boundary and triangles elsewhere. No-slip boundary conditions are imposed at the cylinder surface and Dirichlet boundary conditions $(u,~v) = (1,~0)$ are set at the left, top and bottom edges. 
An outflow boundary condition is defined at the right edge of the domain. 
The initial conditions are such that $(u_0,~v_0) = (0,~0)$. 
As the solution is expected to be smooth, a polynomial order of 7 is used and the time-step is $\Delta t = 0.001$. 

\begin{figure}[h]
\begin{center}
\includegraphics[width=8cm]{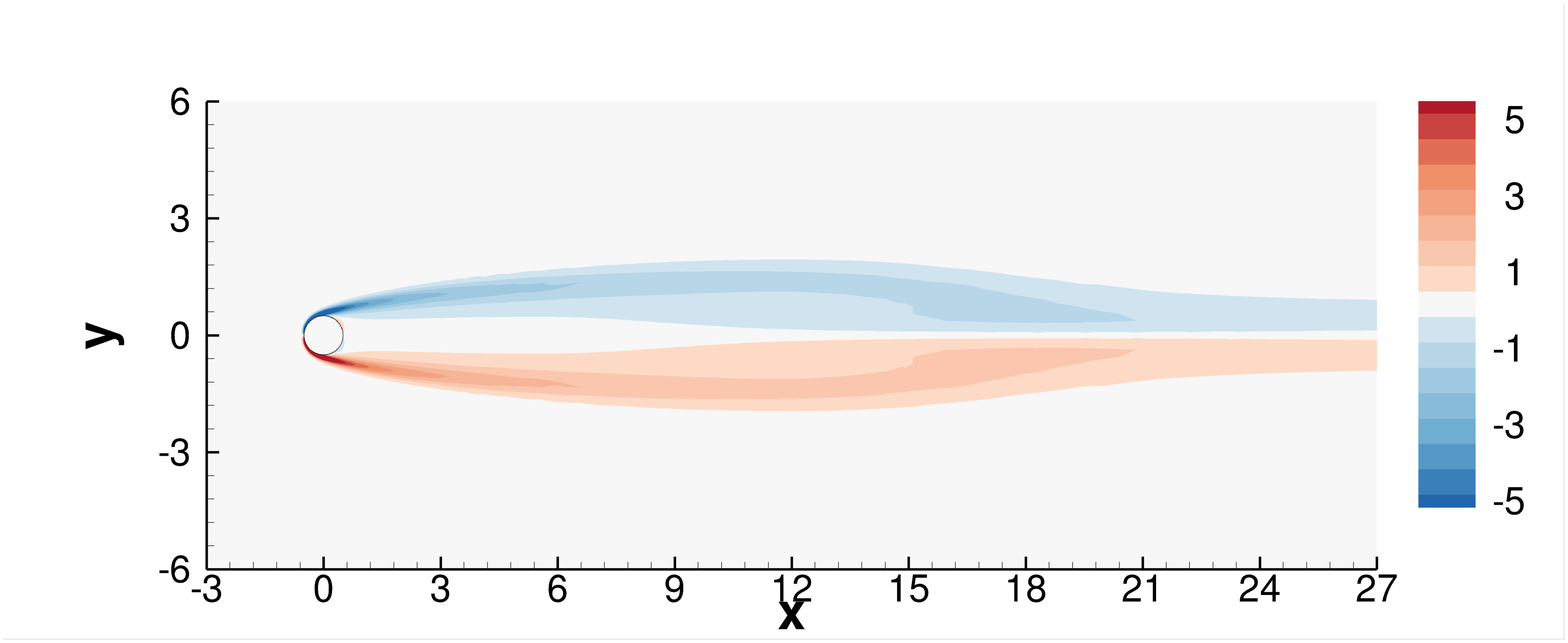}
\caption{\label{Fig_Re300_SS} Vorticity of the  steady-state of the incompressible flow past a two-dimensional cylinder at $Re=300$.}
\end{center}
\end{figure}

The initial parameters are chosen to be $\chi_{\rm{init}} = 1$ and $\Delta_{\rm{init}} = 2$ and the modified Arnoldi iteration method is executed every $T=200$ time units. With these settings, the adaptive SFD method calls the stability analysis method five times and converges in a total of 5622 time units (including 377 Arnoldi steps). The steady-state solution obtained is shown on Fig. \ref{Fig_Re300_SS}.

The last approximation of the dominant eigenvalue (computed using a ``partially converged'' steady-state when $||\bm{\mbox{q}} - \bar{\bm{\mbox{q}} }||_{\rm{inf}}<10^{-2}$) is $\tilde{\sigma}_5 = 0.165$ and $\tilde{f}_5 = 0.493$, which is very similar to the dominant eigenvalue computed using the true steady-state (when $||\bm{\mbox{q}} - \bar{\bm{\mbox{q}} }||_{\rm{inf}}<10^{-8}$) as the base flow, when $\sigma = 0.163$ and $f = 0.470$.

As an extension to Sec. \ref{SubSecRe100}, we showed here that the adaptive SFD method is able to automatically select values of the control coefficient and filter width to control the evolution of the least stable eigenmode of the flow past a cylinder when the Reynolds number is increased. However a long time is necessary to reach convergence.


\subsection{Incompressible flow past an ellipse at $Re=150$ with an angle of attack}
\label{SectionEllipse}

One may argue that if the geometry studied contains axial symmetry (like for the flow past a cylinder), using the SFD method, or any method that provides a steady-state solution of the Navier-Stokes equations, is unnecessary. For example, Mao and Blackburn \cite{mao2014structure} presented a stability analysis (global and local) of the incompressible flow past a square cylinder (up to $Re=300$) using a base flow obtained on a semi-domain with symmetric boundary conditions on the horizontal axis. In this section we apply the adaptive method presented in Sec. \ref{SectionAdaptive} to the incompressible flow past an ellipse at $Re=150$ with an angle of attack of $30^{\circ}$. The major axis of the ellipse is 1 length unit long and the minor axis is 0.4 length unit long. The major axis is used as characteristic length to calculate the Reynolds number. This test case can not be treated with symmetry planes.

The computational domain considered is composed of 856 elements and its dimensions are $-15 \leq x \leq 45$ and $-25 \leq y \leq 25$. The mesh is made of structured curved quadrilaterals close to the cylinder boundary and triangles elsewhere. No-slip boundary conditions are imposed at the ellipse surface and Dirichlet boundary conditions $(u,~v) = (1,~0)$ are set at the left, top and bottom edges. An outflow boundary condition is defined at the right edge of the domain. The initial conditions are such that $(u_0,~v_0) = (0,~0)$. A polynomial order of 7 is used and the time-step is $\Delta t = 0.0025$. 

\begin{figure}[h]
\begin{center}
\includegraphics[width=8cm]{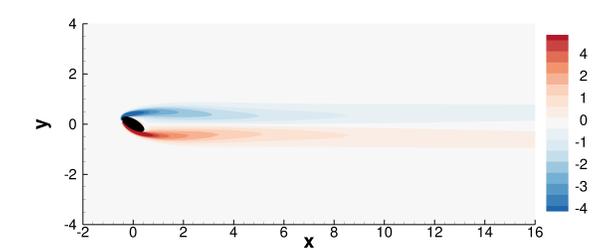}
\caption{\label{Fig_Ellipse} Vorticity of the  steady-state of the incompressible flow past a two-dimensional ellipse at $Re=150$ with an angle of attack of $30^{\circ}$.}
\end{center}
\end{figure}

Nothing is assumed about the stability properties of the flow. We only know that the flow is unstable at this Reynolds number because if DNS simulation is computed, we can observe formation of vortex streets. The initial parameters are randomly chosen to be $\chi_{\rm{init}} = 1$ and $\Delta_{\rm{init}} = 1$ and the modified Arnoldi iteration method is executed every $T=50$ time units. With these settings, the adaptive SFD method calls the stability analysis method twice and converges in a total of 577 time units (including 82 Arnoldi steps). The steady-state solution obtained is shown on Fig. \ref{Fig_Ellipse}.

Note that the second approximation of the dominant eigenvalue (computed using a ``partially converged'' steady-state when $||\bm{\mbox{q}} - \bar{\bm{\mbox{q}} }||_{\rm{inf}}<10^{-2}$) is $\tilde{\sigma}_2 = 0.169$ and $\tilde{f}_2 = 1.287$, which is very similar to the dominant eigenvalue computed using the true steady base flow (when $||\bm{\mbox{q}} - \bar{\bm{\mbox{q}} }||_{\rm{inf}}<10^{-8}$) where $\sigma = 0.168$ and $f = 1.283$. 

This test case shows that an adaptive SFD method can easily converge towards an unstable steady flow with no axial symmetry and without any \textit{a priori} knowledge of the dominant eigenvalue. 


\subsection{Incompressible flow past a rotating cylinder at $Re=100$}
\label{SectionRotatingCyl}

In this section we aim to find the steady-state solution of an other flow that can not be studied with the help of symmetry planes. This test case is the incompressible two-dimensional flow past a rotating cylinder at $Re=100$. The rotation of the cylinder impacts the stability of the flow. As shown by Pralits \textit{et. al}\cite{pralits2010instability}, for a rotation rate $0 \leqslant \alpha \lesssim 1.8$, the flow is unstable and von Kármán vortex streets are present. They become weaker as $\alpha$ increases. This instability is called shedding mode I. If the rotation rate is in the range $1.8 \lesssim \alpha \leqslant 4.85$, the flow becomes stable. If the rotation rate is increased again, a second unstable mode appears (called shedding mode II) for a range $4.85 \leqslant \alpha \leqslant 5.17$. And eventually, for rotation rates above 5.17, the flow is stable. Note that this behaviour is only true in two-dimensions. In three-dimensions the presence of shedding mode I and the range of shedding mode II depend on the spanwise wave number\cite{pralits2013three}. 

Here we are interested in finding a steady-state solution of the unstable mode II, hence we consider a rotation rate $\alpha = 5$. The computational domain considered is composed of 1044 elements and its dimensions are $-15 \leq x \leq 45$ and $-25 \leq y \leq 25$. The mesh is made of structured curved quadrilaterals close to the cylinder boundary and triangles elsewhere. The mesh is fine close to the cylinder boundary because its rotation induces a strong velocity gradient. Also, the region on the right of the cylinder and for $0 \leq y \leq 8$ is refined because that is where the shedding vortices appear when a DNS simulation is executed. Slip boundary conditions are imposed at the cylinder surface such that \textbf{u}$\cdot$\textbf{t}  = $\alpha$ and \textbf{u}$\cdot$\textbf{n}  = $0$,  where \textbf{u} is the velocity vector, \textbf{t} and \textbf{n} are the tangential and normal vectors to the surface respectively. Dirichlet boundary conditions $(u,~v) = (1,~0)$ are set at the left, top and bottom edges. An outflow boundary condition is defined at the right edge of the domain. The initial conditions are such that $(u_0,~v_0) = (0,~0)$. A polynomial order of 7 is used and the time-step is $\Delta t = 0.0005$. 

To compute the adaptive method presented in Sec. \ref{SectionAdaptive}, the initial control coefficient of the SFD method is randomly chosen to be  $\chi_{\rm{init}} = 1$. However some care is taken to the selection of the initial filter width. When DNS simulation is executed, we observe that the frequency of the shedding mode II is very low. Thus the initial filter width of the SFD method must be chosen to be quite high, \textit{e.g.} $\Delta_{\rm{init}} = 5$. Larger filter widths enable us to control instabilities that arise on a larger time scale, but may require an impractically long time to converge. The SFD method is well suited to obtain steady-state solutions of flows with high unstable frequencies. Hence the flow past a cylinder at the unstable shedding mode II is challenging for the SFD method. For this test case the initial filter width is carefully chosen, but nothing is assumed about the dominant eigenvalue of the flow.

\begin{figure}[h]
\begin{center}
\includegraphics[width=8cm]{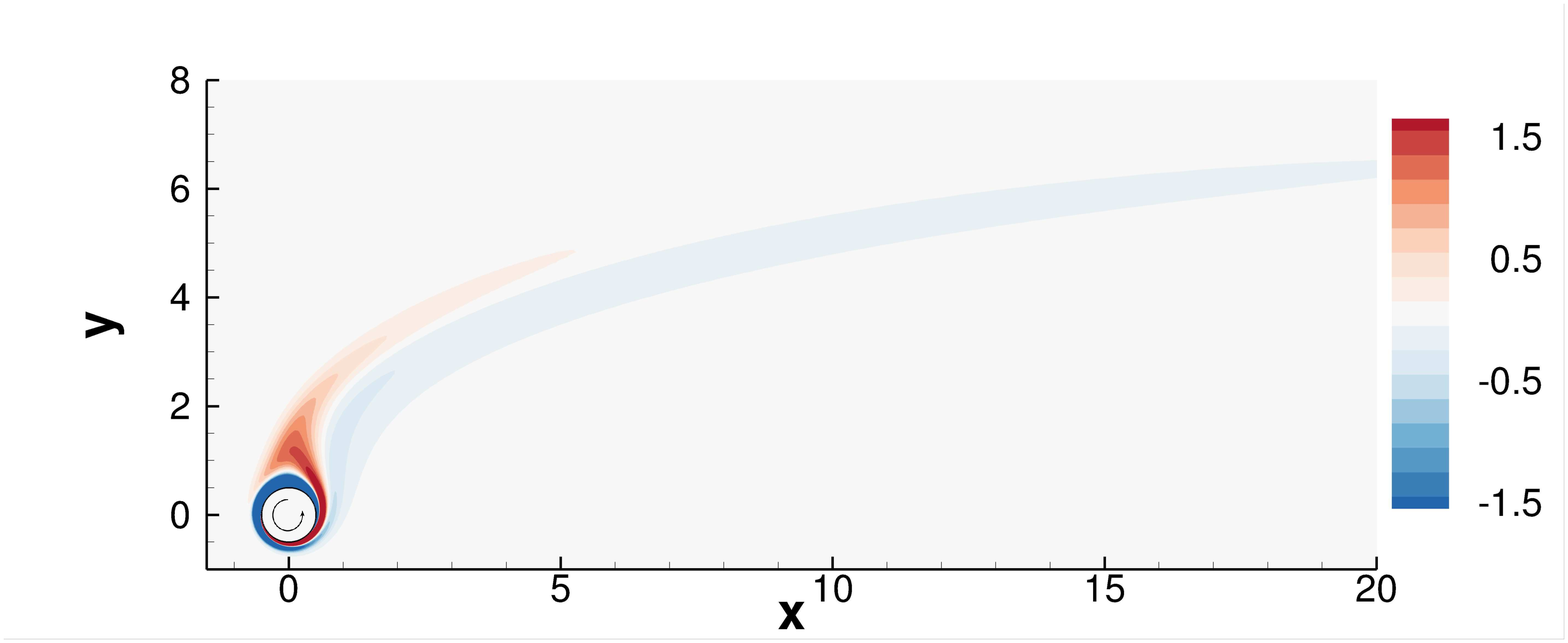}
\caption{\label{Fig_RotCyl} Vorticity of the  steady-state of the incompressible flow past a two-dimensional rotating cylinder at $Re=100$ with a rotation rate $\alpha = 5$.}
\end{center}
\end{figure}

As a long time is necessary for the flow to be established, a relatively large time $T=300$ is chosen. This means that the stability analysis method is executed for the first time after 300 time units. With these settings, the adaptive SFD method calls the stability analysis method twice, and converges in a total of 907 time units (including 145 Arnoldi steps). The steady-state solution obtained is shown on Fig. \ref{Fig_RotCyl}. Note that if a shorter time $T$ is chosen, the stability analysis using a ``partially converged'' base flow does not capture relevant features of the flow. Hence the approximation of the dominant eigenvalue is not good enough and the corresponding SFD parameters only enable a slow convergence towards the steady-state.

The second approximation of the dominant eigenvalue (computed using a ``partially converged'' steady-state when $||\bm{\mbox{q}} - \bar{\bm{\mbox{q}} }||_{\rm{inf}}<10^{-2}$) is $\tilde{\sigma}_2 = 0.039$ and $\tilde{f}_2 = 0.239$, which is very similar to the dominant eigenvalue computed using the true steady-state (when $||\bm{\mbox{q}} - \bar{\bm{\mbox{q}} }||_{\rm{inf}}<10^{-8}$) as the base flow, when $\sigma = 0.036$ and $f = 0.241$. This dominant eigenvalue is similar to the one reported by Pralits \textit{et al}\cite{pralits2010instability}.



%
%
%


\section{Conclusion}

An adaptive procedure to address the issue of selecting appropriate parameters for the SFD method is presented. This algorithm links together a SFD method, a stability analysis method and a one-dimensional model that evaluates the optimum parameters of the SFD method when the dominant eigenvalue is known. This adaptive method is based on several successive computations of the stability analysis method using ``partially converged'' base flows. This approximation is then used to tune the parameters of the SFD method to ensure an optimum convergence towards the steady-state solution.

This adaptive method was successfully applied to obtain unstable steady-states of several flows. The steady-state of the flow past a cylinder was obtained up to the Reynolds number 300. The steady-state of the flow past an ellipse with an angle of attack at $Re=150$ was also presented. This test case illustrates the fact that the adaptive SFD method is an appropriate tool to study flows that do not contain axial symmetry. Finally, the steady-state solution of the unstable mode II of the rotating cylinder was presented. This test case  was fairly challenging for our algorithm but a careful definition of some parameters allows the SFD method to reach an optimum convergence rate. This set of test cases validates our adaptive method, hence we can now use it to investigate the stability of more challenging.

Users of this adaptive SFD method must define several parameters which are the initial control coefficient $\chi_{\rm{init}}$ and filter width $\Delta_{\rm{init}}$, the time $T$ between two consecutive execution of the stability analysis method, the tolerance $\varepsilon_{\mbox{\tiny{Stab}}}$ of the stability analysis method and the tolerance  $\varepsilon_{\mbox{\tiny{Adapt}}}$ such that the SFD parameters are fixed when $||\bm{\mbox{q}} - \bar{\bm{\mbox{q}} }||_{\rm{inf}}<\varepsilon_{\mbox{\tiny{Adapt}}}$. For this study we kept $\varepsilon_{\mbox{\tiny{Stab}}}$ and $\varepsilon_{\mbox{\tiny{Adapt}}}$ constant for all our numerical simulations. The cost of the execution of the modified Arnoldi iteration method is approximately the same as the execution of the SFD method. Hence $\varepsilon_{\mbox{\tiny{Stab}}}$ and $\varepsilon_{\mbox{\tiny{Adapt}}}$ must be selected in order trade-off this cost with an approximation accurate enough of the dominant eigenvalue.

For all our simulations, the steady-state solution was found in several hundreds time units (several thousands for some cases). Even if the convergence rate of the SFD method is optimal, the time (and the computational cost) necessary to reach the steady-state solution may still be important. It may be that the adaptive SFD method could be then used as a global solver to find an initial guess of a Newton's method. This might allow us to design algorithms capable of reaching unstable steady-state solutions in a limited number of iterations.


\section*{Acknowledgments}
The authors would like to thank the Seventh Framework Programme of the
European Commission for their support to the ANADE project (Advances
in Numerical and Analytical tools for DEtached flow prediction) under
grant contract PITN-GA-289428.


\bibliographystyle{unsrt}
\bibliography{Bibliography.bib}

\end{document}